\documentclass[
 aip,
 amsmath,amssymb,
reprint, onecolumn 
]{revtex4-2}

\usepackage{graphicx}
\usepackage{dcolumn}
\usepackage{bm}
\usepackage{color}
\usepackage{bbm}
\usepackage{framed}
\usepackage{amsmath}
\usepackage{amssymb}
\usepackage{mathptmx}
\DeclareMathAlphabet{\mathcal}{OMS}{cmsy}{m}{n} 
\usepackage{subcaption}
\usepackage[utf8]{inputenc}
\usepackage[T1]{fontenc}
\usepackage{etoolbox}
\definecolor{darkgreen}{rgb}{0.05, .65, 0.05}

\definecolor{shadecolor}{rgb}{0.85, .85, 0.85}

\def\bigdlangle{{\big\langle\kern-.23em\big\langle}}
\def\bigdrangle{{\big\rangle\kern-.23em\big\rangle}}
\def\Bigdlangle{{\Big\langle\kern-.35em\Big\langle}}
\def\Bigdrangle{{\Big\rangle\kern-.35em\Big\rangle}}
\def\biggdlangle{{\bigg\langle\kern-.35em\bigg\langle}}
\def\biggdrangle{{\bigg\rangle\kern-.35em\bigg\rangle}}
\def\Biggdlangle{{\Bigg\langle\kern-.35em\Bigg\langle}}
\def\Biggdrangle{{\Bigg\rangle\kern-.35em\Bigg\rangle}}

\def\epsr{\epsilon_\mathrm{r}}
\def\mur{\mu_\mathrm{r}}

\def\linklangle{{\langle\kern-.28em\scalebox{0.35}[1]{$-$}\kern-.18em\langle}}
\def\linkrangle{{\rangle\kern-.18em\scalebox{0.35}[1]{$-$}\kern-.28em\rangle}}
\def\biglinklangle{{\kern.23em\big\langle\kern-.59em\scalebox{0.35}[1]{$-$}\kern-.38em\big\langle\kern.23em}}
\def\biglinkrangle{{\kern.23em\big\rangle\kern-.38em\scalebox{0.35}[1]{$-$}\kern-.59em\big\rangle\kern.23em}}
\def\Biglinklangle{{\kern.0em\Big\langle\kern-.49em\scalebox{0.38}[1]{$-$}\kern-.18em\Big\langle\kern.0em}}
\def\Biglinkrangle{{\kern.0em\Big\rangle\kern-.18em\scalebox{0.38}[1]{$-$}\kern-.49em\Big\rangle\kern.0em}}
\def\bigglinklangle{{\kern.0em\bigg\langle\kern-.62em\scalebox{0.54}[1]{$-$}\kern-.18em\bigg\langle\kern.0em}}
\def\bigglinkrangle{{\kern.0em\bigg\rangle\kern-.18em\scalebox{0.54}[1]{$-$}\kern-.62em\bigg\rangle\kern.0em}}
\def\Bigglinklangle{{\kern.0em\Bigg\langle\kern-.66em\scalebox{0.66}[1]{$-$}\kern-.22em\Bigg\langle\kern.0em}}
\def\Bigglinkrangle{{\kern.0em\Bigg\rangle\kern-.22em\scalebox{0.66}[1]{$-$}\kern-.66em\Bigg\rangle\kern.0em}}

\def\I[#1]#2{\mathbbm{1}^{(#1)}_{#2}}

\def\Tr{\ensuremath{\mathrm{Tr}}}

\def\cT{\ensuremath{\mathcal{T}}}

\def\x0{\ensuremath{\mathbf{x}_\mathrm{0}}}
\def\epsr{\ensuremath{\epsilon_{\mathrm{r}}}}
\def\mur{\ensuremath{\mu_{\mathrm{r}}}}
\def\subEM{_\mathrm{\scriptscriptstyle EM}}
\def\subTE{_\mathrm{\scriptscriptstyle TE}}
\def\subTM{_\mathrm{\scriptscriptstyle TM}}

\def\subCP{_\mathrm{\scriptscriptstyle CP}}

\definecolor{darkblue}{rgb}{0.0,0.0,0.5}
\usepackage[colorlinks,citecolor=darkblue,linkcolor=darkblue]{hyperref}

\makeatletter
\def\@email#1#2{
 \endgroup
 \patchcmd{\titleblock@produce}
  {\frontmatter@RRAPformat}
  {\frontmatter@RRAPformat{\produce@RRAP{*#1\href{mailto:#2}{#2}}}\frontmatter@RRAPformat}
  {}{}
}
\makeatother

\begin{document}

\title{Atoms, Worldlines, and the 
Scalar Approximation}

\author{He Zheng}

\author{Daniel A. Steck}
 \email{pzheng@uoregon.edu, dsteck@uoregon.edu.}
\affiliation{ 
Department of Physics and Oregon Center for Optical, Molecular and Quantum Science,\\
1274 University of Oregon, Eugene, Oregon 97403-1274
}

\date{\today}

\begin{abstract}
The worldline path-integral method for scalar fields has shown promising computational efficiency in evaluating Casimir energies---even in general geometries where some other methods become intractable. However, its current implementation relies on a scalar approximation that treats electromagnetic waves as two independent polarizations. In this work, 
we derive the worldline 
Casimir--Polder potential between $N$ atoms within the
scalar approximation.
Furthermore, we assess the accuracy of this approximation in modeling multiple-atom Casimir--Polder interactions. For two-atom systems, a  sum of the contributions from the two scalar polarizations successfully reproduces the exact force. But when extended to three-body systems, this scalar approach deviates in both magnitude and sign, revealing polarization-mixing effects not captured by the scalar approximation. We explain this by showing how the TE/TM decomposition differs between the worldline method and the Green-tensor formalism. These results reveal limitations of current scalar models. They also motivate the development of methods that more accurately account for the full vector nature of vacuum electromagnetic interactions in general geometries. 

\vspace{2ex}\noindent
\end{abstract}

\maketitle

\section{Introduction}

Dispersion forces are a fundamental class of interactions that arise from quantum mechanical fluctuations in the electromagnetic field. These interactions play a crucial role in multiple-body systems, determining the structure and thermodynamic properties of assemblies of nano-materials, biological systems, and condensed matter. 

Fluctuations in the charge distributions of neutral atoms or molecules lead to the formation of instantaneous dipoles. Dispersion forces arise as a result of these fluctuation-induced interactions, where the dipoles interact with those in neighboring atoms or molecules, generating attractive or repulsive forces. For two-body interactions, the quantum mechanical formulation of this effect is well established through the London dispersion force at short ranges \cite{London1930} or in general the Casimir--Polder force.\cite{CasimirPolder1948} However, in systems with more than two bodies, the interaction energies cannot be reduced to a simple sum of pairwise forces due to the non-additivity of the Casimir--Polder force. Instead, dispersion interactions emerge from the collective interplay of quantum fluctuations across all particles in the system.

A specific class of fluctuation-induced forces arises between atoms and surfaces or between multiple atoms in close proximity to surfaces. These interactions are a generalization of the retarded van der Waals forces \cite{Dzyaloshinskii1961} and are relevant in experimental setups where the interacting particles are separated by distances typically up to around 10~$\mu$m.
Recent theoretical and experimental investigations have demonstrated that Casimir--Polder interactions---which tend to be attractive---can be tuned in strength through modifications to experimental parameters.\cite{Venkataram2020} In specific cases, the sign of the interaction can also be altered, typically by adjusting the dielectric properties of the background \cite{Munday2009} or employing carefully designed geometries that allow for repulsive forces, though such repulsive geometries are often sensitive to small perturbations and inherently unstable.\cite{Levin2010} This tunability opens new avenues for exploring quantum field theory phenomena and condensed matter systems, enabling the control of quantum forces in modern experimental setups and quantum-inspired devices.\cite{Folman2002, Fortagh2007} Casimir--Polder forces also play a pivotal role in the manipulation of ultra-cold atoms near surfaces and in the design of micro- and nano-structures, where precision control of atom-surface interactions is critical.\cite{Diedrich1989, Meekhof1996, Pitaevskii2003, Antezza2004} Moreover, recent developments have highlighted that thermal fluctuations and out-of-equilibrium conditions can significantly modify the strength and behavior of these forces.\cite{Obrecht2007, Antezza2008, Laliotis_2014}

In the simplest theoretical framework, dispersion interactions are approximated as the sum of pairwise interactions between atoms. While this approach can be adequate for dilute systems, when many-body effects are weak, it fails to capture significant collective interactions that become pronounced in dense or strongly interacting systems, where the pairwise approximation introduces substantial errors. In computational chemistry, for example, Density Functional Theory (DFT) is sometimes enhanced with empirical van der Waals corrections to account for dispersion forces.\cite{Grimme2006} These corrections are typically pairwise terms and based on parameterized formulas. However, pairwise summation approaches neglect the non-additivity that arises from collective quantum fluctuations involving more than two particles. This leads to considerable errors especially in dense or strongly interacting systems where many-body effects are significant. An advanced approach based on quantum electrodynamics provides a more rigorous framework for calculating dispersion interactions from fundamental principles.\cite{PhysRevA.25.2473, Power1985, Power1980} The idea is to compute the energy shift of the ground state of the system using perturbation theory, specifically due to the dipole coupling between the atom(s) and the field. Analytic QED-based calculations typically require knowledge of the photon mode functions that describe how the field behaves in the presence of specific boundary conditions, which are only easily solvable in highly symmetric cases. When the system is more complex or lacks symmetry, solving for the mode functions becomes 
intractable. This limits the applicability of these methods to simpler scenarios where the boundary conditions are well-understood.

A challenge in calculating dispersion interactions lies in modeling electromagnetic fields in complex geometries. Scattering-based approaches \cite{Johnson2011, reid2009, reid_2013} provide a rigorous framework by modeling virtual photon-mediated interactions and capturing polarization coupling through multiple scattering processes. Recent advancements, including boundary element methods and spectral decomposition techniques, have improved the efficiency of solving Maxwell’s equations in arbitrary geometries. Open-source tools like SCUFF-EM \cite{SCUFFEM} have further made these methods accessible for precise modeling of Casimir and Casimir--Polder forces.

The worldline path-integral method offers an alternative framework for numerically calculating Casimir energies.\cite{Gies2003, Gies2006, Gies2006a, Gies2006b, Gies2006c, Klingmuller2008, Schafer2016, Schafer2012, Weber2009, Weber2010, Fosco2010, FranchinoVinas20}  The method is nonperturbative, in the sense that, for example, it handles arbitrarily large dielectric permittivities. This approach generates an ensemble of closed, stochastic trajectories and evaluates the interaction energy by averaging over the material properties of the interacting bodies. A key advantage of the worldline approach is its ability to efficiently handle arbitrary geometries and boundary conditions. However, while promising, the method has primarily been applied to simpler geometries and systems involving scalar fields, where the polarization-decomposition approximation simplifies full electromagnetism.

The scalar approach to Casimir physics has proven particularly fruitful in the literature. For example, a semi-classical approximation based on geometric optics has shown that scalar models can reproduce correct results for symmetric configurations, such as parallel plates, and provide useful insights into more complex setups through the proximity force approximation.\cite{Scardicchio_2005, Jaffe_2004, Schoger2022} Other seminal work on S-matrix-based approaches provides direct tests of the scalar approximation from a field-theoretic perspective.\cite{PhysRevD.72.021301, Jaffe:2003bx} While these studies demonstrate that the scalar framework can accurately capture Casimir forces in many contexts, its validity remains uncertain 
in more general 
scenarios---especially when polarization mixing is pronounced. In this paper, we identify and analyze such a system to critically assess the limitations of the scalar approximation.

Recent developments in the worldline method have focused on incorporating the coupling between the field and the dielectric or magnetic properties of the media.\cite{Mackrory2016, Zheng2026} These studies have so far assumed symmetric geometries, allowing the fields to be decomposed into, for example, uncoupled transverse-electric (TE) and transverse-magnetic (TM) polarization modes, each treated as an independent scalar field. On the computational side, Ref.~[\onlinecite{Mackrory2024}] demonstrates efficient partial-averaging techniques for differentiation of scalar worldline path integrals in atom–plane and plane–plane configurations. To fully exploit the computational potential of the worldline approach in general geometries, it is essential to extend the method to account for the full vectorial nature of the electromagnetic field, thereby capturing polarization mixing and other effects that arise beyond idealized symmetric setups.

This paper is organized as follows: First, we provide an overview of the multi-particle scattering theory of dispersion forces and derive the interaction potentials with Green tensors of two and three atoms. In the following section, we develop path integrals for general Casimir energies in the scalar representation of electromagnetic waves. In this formalism, the retarded atom–atom Casimir--Polder potential is obtained by introducing localized perturbations in magneto-dielectric properties that model the presence of neutral atoms. We then generalize the worldline results for each polarization mode to discrete systems with arbitrary $N$ atoms and calculate explicitly up to three-body interaction potentials. Finally, we compare these results with those from the Green tensor scattering approach, which employs the free-space Green function in a plane-wave decomposition to describe the EM field response due to multiple dipole sources. 

\section{Scattering Theory of Dispersion Interactions}

Within a quantum electrodynamical framework, consider a collection of $N$ neutral, nonpolar atoms (or molecules) in vacuum, each located at $\mathbf{r}_i$. These particles possess instantaneous dipoles from linear response to the electromagnetic field. Dispersion interactions arise from the coupling of the dipoles with the field, described by the interaction Hamiltonian 
\begin{align}
\hat{H}_{\mathrm{int}} =-\sum_{i=1}^N \hat{\mathbf{d}}_{i} \cdot\hat{\mathbf{E}}\bigl(\mathbf{r}_{i}\bigr),
\end{align}
where $\hat{\mathbf{d}}_{i}$ is the electric dipole operator for the $i$th particle and $\hat{\mathbf{E}}$ is the quantized electric field operator. In such a multi-dipole system, the local field at the $n$th atom's position comprises both the ``bare" vacuum fields $\hat{\mathbf{E}}_{0}$ and the induced fields from all the dipole sources.\cite{Rosa_2011} Symbolically,
\begin{align}
\hat{\mathbf{E}}(\mathbf{r}_n,\omega)=\hat{\mathbf{E}}_{0}(\mathbf{r}_n,\omega)+\sum_{m=1}^N \mathbf{G}^{(0)}(\mathbf{r}_n,\mathbf{r}_m,\omega) \cdot \hat{\mathbf{d}}_m(\omega),
\label{totalEfield}
\end{align}
where $\mathbf{G}^{(0)}$ is the free dyadic Green function encoding propagation of the field from atom $m$ to atom $n$, which satisfies
\begin{align}
\Big[\nabla \times \nabla \times {}-{}\frac{\omega^2}{c^2}\Big] \mathbf{G}^{(0)}(\mathbf{r}, \mathbf{r'}, \omega) = \mu_0 \omega^2 \delta^3 (\mathbf{r}-\mathbf{r'}).
\end{align}
That is, the free-space Green tensor $\mathbf{G^{\bm{(0)}}}(\mathbf{r}, \mathbf{r'}, \omega)$ is defined as the solution to the impulse-driven wave equation in response to a point source at $\mathbf{r'}$.

The 
Green-tensor formalism underlies a standard derivation of dispersion interactions. 
In this derivation, we can use Eq.~(\ref{totalEfield})
in the equation $\mathbf{d}_n(\omega)=\mathbf{\alpha}_n(\omega)\cdot\mathbf{E}(\omega,\mathbf{r}_n)$, where $\mathbf{\alpha}_n$ is the polarizability
of the $n$th atom, leading to its induced dipole moment $\mathbf{d}_n$.
Then we solve the result for the collective dipole moment, 
and read off the total polarizability as the
coefficient of $\mathbf{E}_0$. 
Putting this polarizability into the well-known expression for
the Casimir--Polder interaction,\cite{McLachlan63a, McLachlan63b, Wylie84}
\begin{equation}
  V\subCP=-\frac{\hbar}{2\pi}\mathrm{Im} \int_0^\infty d\omega\,
  \mathrm{Tr}\big[\mathbf{\alpha}(\omega)\cdot G^{(\mathrm{s})}(\mathbf{r},\mathbf{r},\omega)\big]
\end{equation}
where $G^{(\mathrm{s})}$ is the scattered part of the Green tensor
(i.e., the Green tensor with $G^{(0)}$ subtracted away),
we can use the Hellmann--Feynman theorem\cite{Feynman39}
to compute the induced interaction between the dipoles.
This approach gives the Casimir--Polder interaction of a collection of $N$ atoms in terms of the free-space Green tensors $\mathbf{G}^{(0)}(\omega)$, where the Taylor expansion of the $\log$ produces terms that correspond to one-body, two-body, three-body interactions, and so on:\cite{Rosa_2011}
\begin{align}
V\subCP&=\frac{\hbar}{2\pi} \mathrm{Im} \int_0^{\infty} \!\! d\omega \, \Tr \log \Big [1
- \boldsymbol{\alpha}(\omega) \cdot \mathbf{G}^{(0)}(\omega) \Big ] \nonumber\\
&=-\frac{\hbar}{2\pi} \mathrm{Im} \int_0^{\infty} \!\! d\omega \, \Tr \Big[ \boldsymbol{\alpha}(\omega) \cdot \mathbf{G}^{(0)}(\omega) \nonumber\\
&\hspace{5mm}+\frac{1}{2} \boldsymbol{\alpha}(\omega) \cdot \mathbf{G}^{(0)}(\omega) \cdot \boldsymbol{\alpha}(\omega) \cdot \mathbf{G}^{(0)}(\omega) \nonumber\\
&\hspace{5mm}+\frac{1}{3} \boldsymbol{\alpha}(\omega) \cdot \mathbf{G}^{(0)}(\omega) \cdot \boldsymbol{\alpha}(\omega) \cdot \mathbf{G}^{(0)}(\omega) \cdot \boldsymbol{\alpha}(\omega) \cdot \mathbf{G}^{(0)}(\omega) \nonumber\\
&\hspace{5mm}+O\big(\boldsymbol{\alpha}^4 \big)\Big].
\label{eq:trlogG}
\end{align}
Here, $\boldsymbol{\alpha}(\omega)$ represents the atomic polarizability tensor.
However, the symbols $\boldsymbol{\alpha}$ and $\mathbf{G}^{(0)}$
in Eq.~(\ref{eq:trlogG}) are not the polarizability of a single
atom and this Green tensor (each of which is a $3 \times 3$ object), respectively; rather, they are
$3N\times 3N$ tensors representing the composite quantities over the $N$ atoms. Additionally the trace in Eq.~(\ref{eq:trlogG}) sums over both the polarization degrees of freedom and the locations of the atoms.
The perturbative expansion is justified when the product $\boldsymbol{\alpha}(\omega) \cdot \mathbf{G}^{(0)}(\omega)$ is sufficiently small, which is typically the case here. This expression reveals the nature of the interaction as a scattering problem of electromagnetic waves in vacuum with multiple dipole scatterers. For simplicity, we consider ground-state, spherically symmetric atoms with scalar polarizability $\alpha_0$ in the static field limit. The contribution from the first term is a sum of single-particle self energies, which are formally divergent and must be renormalized away. The second term contributes the sum of pairwise retarded van der Waals interactions, with self-energy $m=n$ excluded:
\begin{align}
V_2=-\frac{\hbar}{2\pi} \mathrm{Im} \int_0^{\infty}\!\! d\omega \, \frac{1}{2} \sum_{m \neq n} &\alpha_0^2 \, G_{ij}^{(0)}(\mathbf{r}_n, \mathbf{r}_m,\omega)\, G_{ji}^{(0)}(\mathbf{r}_m, \mathbf{r}_n,\omega).
\label{eq:v2}
\end{align}
Here we assume the Einstein convention of summing over repeated indices, and we are writing out the spatial part of the trace.
Higher-order terms in the series contribute multiple-body corrections beyond the pairwise sum to the total interaction potential.

In practice, the free-space Green tensor can be expanded in terms of a spectrum of monochromatic plane waves propagating in the $\mathbf{k}$ direction, with $k = \omega/c$. We decompose $\mathbf{k}$ into transverse components $(k_x, k_y)$ and a longitudinal component $k_z = \smash{(k^2 - k_x^2 - k_y^2)^{1/2}}$, yielding the expression of the Green tensor in the plane-wave decomposition,
\begin{align}
G_{ij}^{(0)}(\mathbf{r},0,\omega)&=\frac{i}{8\pi^2 \epsilon_0} \int_{-\infty}^{\infty} \!\! dk_x \int_{-\infty}^{\infty} \!\! dk_y \frac{1}{k_z} (k^2 \delta_{ij}+\partial_i \partial_j) \, e^{i(k_x x+k_y y+k_z |z|)},
\end{align}
where the source is located at the origin. With $\mathbf{k}$ as the direction of propagation, the projection operator acting on the complex exponential can be expressed as outer products of TE/TM unit vectors as defined relative to $\hat{z}$, 
\begin{align}
G_{ij}^{(0)}(\mathbf{r},0,\omega)&=\frac{i}{8\pi^2 \epsilon_0} \int_{-\infty}^{\infty} \!\! dk_x \int_{-\infty}^{\infty}\!\! dk_y \, \frac{k^2}{k_z} (\hat{e}_i \hat{e}_j+\hat{h}_i \hat{h}_j) \, e^{i(k_x x+k_y y+k_z |z|)},
\end{align}
where the TE and TM unit vectors are respectively defined as
\begin{align}
\hat{e} &= \frac{k_y}{k_\rho} \hat{x} - \frac{k_x}{k_\rho} \hat{y} = \frac{\hat{k} \times \hat{z}}{|\hat{k} \times \hat{z}|} \nonumber\\
\hat{h} &= -\frac{k_z k_x}{k k_\rho} \hat{x} - \frac{k_z k_y}{k k_\rho} \hat{y} + \frac{k_\rho}{k} \hat{z} = \hat{e} \times \hat{k},
\label{eq:unitvector}
\end{align}
and $k_\rho=\smash{(k_x^2 + k_y^2)^{1/2}}$ denotes the magnitude of the transverse component of the wave vector. The Green tensor now includes contributions from both TE and TM modes [where the electric (TE)
or magnetic (TM) field is perpendicular to the $z$-axis].

When $N=2$, only one pair of atoms is present, and the interaction energy contains exactly one term,
\begin{align}
V\subCP&=-\frac{\hbar \alpha_1 \alpha_2}{2\pi}\!\! \int_0^{\infty}\!\! ds \, \Tr \Big [ \mathbf{G}^{(0)}(\mathbf{r}_1, \mathbf{r}_2,is)\! \cdot \!\mathbf{G}^{(0)}(\mathbf{r}_2, \mathbf{r}_1,is) \Big ] \nonumber\\
&=-\frac{\hbar \alpha_1 \alpha_2}{2\pi} \big(\frac{1}{8\pi^2 \epsilon_0}\big)^2 \int_0^{\infty} \!\! ds \, \int_{-\infty}^{\infty} \!\! dk_x dk_y dk_x' dk_y' \nonumber\\
&\hspace{5mm}\times \frac{s^4/c^4}{\kappa(s,k_x,k_y)\kappa(s,k_x',k_y')} (\hat{e}_a \hat{e}_b + \hat{h}_a \hat{h}_b)(\hat{e}'_b \hat{e}'_a + \hat{h}'_b \hat{h}'_a) \nonumber\\
&\hspace{5mm}\times e^{i(k_x (x_1-x_2)+k_y (y_1-y_2))}e^{-\kappa(s,k_x,k_y) |z_1-z_2|}\nonumber\\
&\hspace{5mm}\times e^{i(k_x' (x_2-x_1)+k_y' (y_2-y_1))}e^{-\kappa(s,k_x',k_y') |z_2-z_1|}, 
\label{eq:vtensor}
\end{align}
where the wave vectors satisfy
\begin{align}
k&=is/c \nonumber\\
k_z&=i\sqrt{s^2/c^2+k_x^2+k_y^2}\equiv i\kappa(s,k_x,k_y).
\end{align}
Note that this imaginary-frequency form is equivalent, via Wick rotation, to the result in Eq.~(\ref{eq:trlogG}). Using the definitions of the TE/TM vectors in Eq.~(\ref{eq:unitvector}), the tensor contraction part of this expression is given by
\begin{equation}
[\hat{e} \hat{e}] = 
\left(\begin{array}{rcl}
\frac{k_y^2}{k_\rho^2}&\frac{-k_x k_y}{k_\rho^2}&0\\ \frac{-k_x k_y}{k_\rho^2}&\frac{k_x^2}{k_\rho^2}&0\\0&0&0 
\end{array}\right)
=: [\mathbf{A}_e]
\end{equation}
and
\begin{equation}
[\hat{h} \hat{h}] = 
\left(\begin{array}{rcl}
\frac{k_x^2 k_z^2}{k_\rho^2 k^2}&\frac{k_x k_y k_z^2}{k_\rho^2 k^2}&\frac{-k_x k_z}{k^2}\\ \frac{k_x k_y k_z^2}{k_\rho^2 k^2}&\frac{k_y^2 k_z^2}{k_\rho^2 k^2}&\frac{-k_y k_z}{k^2}\\ \frac{-k_x k_z}{k^2}&\frac{-k_y k_z}{k^2}&\frac{k_\rho^2}{k^2} \
\end{array}\right)
=: [\mathbf{A}_h],
\end{equation}
which represent TE and TM contributions, respectively. Note that the tensor contraction in Eq.~(\ref{eq:vtensor}) corresponds to tracing the matrix products, with each term given by
\begin{align}
\mathrm{Tr} (\mathbf{A}_e \mathbf{A}_e'^{\mathrm{T}}) &=\frac{(k_x k_x'+k_y k_y')^2}{k_\rho'^2 k_\rho^2} \nonumber\\
\mathrm{Tr} (\mathbf{A}_h \mathbf{A}_h'^{\mathrm{T}})&=\frac{(k_x k_x' k_z k_z'+k_x^2 k_\rho'^2+k_y k_y' k_z k_z'+k_y^2 k_\rho'^2)^2}{k'^2 k_\rho^2 k^2 k_\rho'^2}\nonumber\\
\mathrm{Tr} (\mathbf{A}_h \mathbf{A}_e'^{\mathrm{T}})&=\frac{(k_x' k_y-k_x k_y')^2 k_z^2}{k_\rho^2 k_\rho'^2 k^2}\nonumber\\
\mathrm{Tr} (\mathbf{A}_e \mathbf{A}_h'^{\mathrm{T}})&=\frac{(k_x' k_y-k_x k_y')^2 k_z'^2}{k'^2 k_\rho^2 k_\rho'^2}.
\label{eq:tracing}
\end{align}
The first two terms are pure TE/TM components and the last two terms characterize the mixing between these modes. To further simplify, we assume the two atoms are separated along the $z$-axis, so $x_1=x_2=y_1=y_2=0$. By explicitly integrating all $k$'s and the imaginary frequency $s$, we combine each contribution to obtain
\begin{align}
&V^{\mathrm{TE}}+V^{\mathrm{TM}} + 2V^{\mathrm{TE/TM}} \nonumber\\
&=\Big(-\frac{3\hbar c \alpha_1 \alpha_2}{16\pi}-\frac{73 \hbar c \alpha_1 \alpha_2}{16\pi}-\frac{\hbar c \alpha_1 \alpha_2}{\pi}\Big) \frac{1}{(4\pi \epsilon_0)^2 z^7}\nonumber\\
&=-\frac{23 \hbar c \alpha_1 \alpha_2}{4\pi} \frac{1}{(4\pi \epsilon_0)^2 z^7}.
\label{eq:component}
\end{align}
This result agrees with the full potential obtained from different methods; see for example Refs.~\onlinecite{Power1985, CasimirPolder1948}. 

The calculation above directly generalizes to $N=3$. We present the details of this case in the Appendix. 
Suppose three atoms are located along the $z$-axis, with positions $z_1$, $z_2$, $z_3$ and equally spaced by $R/2$, forming a collinear geometry. The energy of this configuration involves the two-body contributions from every possible pairing and a three-body cross energy proportional to $\alpha_1 \alpha_2 \alpha_3$. The latter in terms of the Green tensors is given by
\begin{align}
V\subCP(z_1,z_2,z_3)&= -\frac{\hbar}{2\pi} \int_0^\infty \!\! ds \, \mathrm{Tr} \Big [\alpha_1 \mathbf{G}^{(0)}_{12} \cdot \alpha_2 \mathbf{G}^{(0)}_{23} \cdot \alpha_3 \mathbf{G}^{(0)}_{31} +\alpha_1 \mathbf{G}^{(0)}_{13} \cdot \alpha_2 \mathbf{G}^{(0)}_{21} \cdot \alpha_3 \mathbf{G}^{(0)}_{32} \Big].
\label{eq:3bd-pot}
\end{align}
Here, the three-body cross interaction originally involves $3! = 6$ combinatoric terms arising from all possible permutations of the interactions among three atoms. However, due to the cyclic symmetry of the trace operation and Green tensors, these reduce to two unique terms. The factor 1/3 that appears explicitly in Eq.~(\ref{eq:trlogG}) is effectively incorporated by accounting for this implicit symmetry. The evaluation of Eq.~(\ref{eq:3bd-pot}) eventually leads to 
\begin{align}
V\subCP(R) = -186 \frac{\hbar c \alpha_1 \alpha_2 \alpha_3}{\pi} \frac{1}{(4\pi \epsilon_0)^3 R^{10}}.
\label{eq:colinear_correct}
\end{align} 
Again, this result is consistent with the standard expression.\cite{Power1985} The $N=3$ case has richer non-trivial configurations compared to $N=2$. We examine another example---three atoms forming an equilateral triangle with side length $R$. In this case, we obtain
\begin{align}
V\subCP(R) = + 5.2 \frac{\hbar c \alpha_1 \alpha_2 \alpha_3}{\pi} \frac{1}{(4\pi \epsilon_0)^3 R^{10}}.
\label{eq:triangle_correct}
\end{align} 
While the two-body Casimir--Polder interaction is always attractive, the three-body force can be either attractive or repulsive depending on the specific geometry. Using the full tensor Green functions in free space, this approach includes all necessary polarization couplings and multi-atom effects.

\section{Worldline Path Integrals}

The worldline approach to quantum field theory formulates quantum processes in terms of trajectories through spacetime (i.e., worldlines). 
It is thus a first-quantized formulation, based on a second-order treatment of QED.
Physical quantities, such as effective actions or heat kernels,
can be expressed as path integrals over these trajectories.
\cite{Schubert2001, Corradini2015, Edwards19, Schmidt94}
The worldline method
is especially valuable as a numerical technique, and there has
recently been a renewed interest in this area, for example
in nonperturbative quantum field theory,\cite{FranchinoVinas19}
curved space,\cite{Corradini20}
and improved sampling methods.\cite{Ahumada23}
There has also been recent analytic work on 
heat kernels for semitransparent boundary conditions.\cite{FranchinoVinas22}

In particular, the worldline approach facilitates the numerical computation of Casimir energies. It proves to be effective in scalar representations of electromagnetism. In wave propagation problems, scalar representations can effectively approximate vector field behavior in the paraxial regime, particularly when the field’s polarization remains stable during propagation. This method starts by assuming planar symmetry, and the wave decouples into two independent scalar fields for the electric and magnetic field, namely TE and TM modes. The equation of motion for each polarization mode defines a Green operator, which is then used to calculate the propagator and the field partition function, followed by the free energy of the system. 

In this section, we develop the general theoretical framework for the Casimir--Polder potential within the worldline formalism. We show that the two-body energy is calculated accurately within the scalar approximation of the worldline method. We then specialize to 
three-atom energies
to show that the TE/TM decomposition is not sufficient to capture the full dispersion interaction potential of an $N$-body discrete system.

\subsection{Transverse Electric Mode}
The unrenormalized ground-state Casimir energy of either the TE or TM scalar mode coupled to a magneto-dielectric background in $D$ spacetime dimensions is given by the worldline expression \cite{Mackrory2016}
\begin{align}
E_{\subEM} &= -\frac{\hbar c}{2(2\pi)^{D/2}} \int_0^\infty \!\! \frac{d\cT}{\cT^{1+D/2}} \, \int d^{D-1} x_0 \,
\biggdlangle \langle \epsr \mur \rangle_{\mathbf{x}(\tau)}^{-1/2} e^{-\cT \langle V \rangle_{\mathbf{x}(\tau)}} \biggdrangle_{\mathbf{x}(\tau)},
\label{eq:generalcas}
\end{align}
where the double angle brackets represent an ensemble average over closed Wiener paths that originate at $\mathbf{x}_0$ with proper time $\cT$ (for example, the double angle brackets encompass the Gaussian weight of the Wiener paths). 
Here $\epsr(\mathbf{r})$ represents the relative dielectric permittivity $\epsilon(\mathbf{r})/\epsilon_0$, and $\mur(\mathbf{r})$ is the relative magnetic permeability $\mu(\mathbf{r})/\mu_0$.
The matter-induced potential $V$ in the exponential takes different forms depending on the polarization mode:
 \begin{align}
V\subTE = \frac{1}{2} [\,(\nabla \log \sqrt{\mur})^2-\nabla^2 \log \sqrt{\mur} \,] \nonumber\\
V\subTM = \frac{1}{2} [\,(\nabla \log \sqrt{\epsr})^2-\nabla^2 \log \sqrt{\epsr} \,].
\label{eq:potential}
\end{align}
Here, the position-dependent permittivity $\epsilon(\mathbf{r})$ and permeability $\mu(\mathbf{r})$ 
represent the magnetodielectric bodies that interact via
the Casimir effect.  For example, two (planar) dielectric half-spaces
separated by distance $d$ could be represented by $\epsilon(z\leq 0)=\epsilon(z\geq d)=\epsilon'$ for some
constant value $\epsilon'$ of the permittivity of the half-spaces,
and $\epsilon(0\leq z\leq d)=\epsilon_0$ for the gap in between.
(Note that Casimir energies can be computed in terms of boundary
conditions or in terms of magnetodielectric properties of media;
here, we do the latter.)

The Casimir--Polder potential is defined as the change in energy resulting from bringing an atom in proximity to magnetodielectric bodies. Ignoring dispersion, the potential as a function of the atomic position $\mathbf{r}$ can be obtained via the first-order variation of the Casimir energy with respect to the small dielectric and magnetic perturbations introduced by the atom. This is formally expressed as
\begin{align}
V_{\subCP} (\mathbf{r}) = \frac{\alpha_0}{\epsilon_0} \Big (\frac{\delta E_{\subEM}}{\delta \epsr} \Big )+\beta_0 \mu_0 \Big (\frac{\delta E_{\subEM}}{\delta \mur} \Big ),
\label{eq:vcp}
\end{align}
where $\alpha_0$ is the static polarizability of the atom and $\beta_0$ is the static magnetizability. For TE polarization, the (unrenormalized) Casimir--Polder potential is given by
\begin{align}
V_{\subCP}^{\subTE} (\mathbf{r})=\frac{\hbar c \alpha_0}{4(2\pi)^{D/2} \epsilon_0} \int_{0}^{\infty} \!\! \frac{d\cT}{\cT^{1+D/2}} \,\biggdlangle \langle \epsilon_{r} \rangle^{-3/2}_{\mathbf{x_r}(\tau)} \biggdrangle_{\mathbf{x_r}(\tau)},
\label{eq:vte_final}
\end{align}
assuming a nonmagnetic atom interacts with nonmagnetic media. Here, the path $\mathbf{x_r}(\tau)$ represents a Brownian bridge pinned so that $\mathbf{x_r}(0)=\mathbf{x_r}(\cT)=\mathbf{r}$.
Given the spatial dependence of the dielectric permittivity, the Casimir--Polder potential can be numerically evaluated using Monte Carlo methods based on Eq.~(\ref{eq:vte_final}). While this is a simplified model that suffices for our comparison of polarization decomposition with the Green tensor method, a discussion of more realistic conditions is provided in, e.g., Ref.~[\onlinecite{Babb2004}].

The one-atom scenario serves as a basis to develop the multiple-body theory in the worldline formalism. Consider the case of adding a second atom near the dielectric medium, where now the two atoms and the medium interact with each other. In addition to the contribution from Eq.~(\ref{eq:vte_final}), we are interested in the $\alpha^2$ contributions to $V_{\subCP}$, corresponding to the second functional derivatives of the Casimir energy. 

Similar to the one-body case, the presence of a second atom introduces another dielectric delta function that perturbs the vacuum. For two atoms located at $\mathbf{r}_1$ and $\mathbf{r}_2$, the final form of the TE potential is given by
\begin{align}
V_{\subCP} (\mathbf{r}_1,\mathbf{r}_2) &= \frac{-3 \hbar c \alpha_1 \alpha_2}{8 (2\pi)^{D/2} \epsilon_0^2} \int_{0}^{\infty} \!\!\frac{d\cT}{\cT^{2+D/2}} \,\int_0^\cT \!\! d\tau' \, \frac{e^{-r^2/2\tau'(1-\tau'/\cT)}}{[2\pi \tau'(1-\tau'/\cT)]^{(D-1)/2}} \biggdlangle \langle \epsilon_{r} \rangle^{-5/2}_{\mathbf{x}(\tau)} \biggdrangle_{\mathbf{x}(\tau)|\mathbf{x}(0) = \mathbf{x}(\cT) = \mathbf{r}_1, \mathbf{x}(\tau')=\mathbf{r}_2}.
\label{eq:te2atoms}
\end{align}
The exponential factor is then the probability density function of such a Brownian bridge to cross the second atom's position. This probability density is a Gaussian distribution with zero mean and variance $\smash{\tau'(1-\tau'/\cT)}$ and arises from the path-pinning statistics in stochastic calculus. 
(For an application of similar pinned path integrals to the numerical calculation of propagators, see Ref.~[\onlinecite{Edwards22}].)

In vacuum, the potential from Eq.~(\ref{eq:te2atoms}) can be analytically evaluated in the absence of the macroscopic media ($\epsr=1$). In a 3+1 dimensional space-time, the expression reduces to
\begin{align}
V_{\subCP}^{\subTE} (r) = -\frac{3 \hbar c \alpha_1 \alpha_2}{8 \pi (4\pi \epsilon_0)^2} \frac{1}{r^7}.
\label{eq:te_pot}
\end{align}
This is the TE component of the Casimir--Polder potential between two neutral atoms from the worldline calculations. The coefficient for the full long-range potential is 23/4, so this TE-component calculation contributes about 6.5\% to the full force. However, notice that this coefficient is twice as large compared to the corresponding Green tensor's TE contribution Eq.~(\ref{eq:component}). 

It is straightforward to extend the above expressions to the $k$th-order Casimir--Polder potential for $N$ atoms located at positions $\mathbf{r}_1, \dots, \mathbf{r}_N$ (for $k\leq N$). The electromagnetic field fluctuations are treated as paths conditioned on specific configurations of the atoms
\begin{align}
V_{\subCP}^{\subTE(k)}(\mathbf{r}_1, &\dots, \mathbf{r}_N) = (-1)^{k+1} \frac{(2k-1)!! \hbar c}{2^{k+1} k! (2\pi)^{D/2} \epsilon_0^k} \sum_{k_1,\dots,k_k=1}^{N} \alpha_{k_1} \dots \alpha_{k_k} \int_0^\infty \!\! \frac{d\cT}{\cT^{k+D/2}} \, \int_0^\cT \prod_{i=2}^k d\tau_i \nonumber \\
    &\quad \times f_{\mathbf{x}(\tau_2)}(\mathbf{r}_2) f_{\mathbf{x}(\tau_3)}(\mathbf{r}_3) \dots f_{\mathbf{x}(\tau_k)}(\mathbf{r}_k) \, \biggdlangle \langle \epsr \rangle ^{-k-1/2}_{\mathbf{x}(\tau)} \biggdrangle_{\mathbf{x}(\tau) \mid \mathbf{x}(0)=\mathbf{r}_{k_1}, \mathbf{x}(\tau_j)=\mathbf{r}_{k_j}, \mathbf{x}(\cT)=\mathbf{x}(0)},
\label{eq:multi_te1}
\end{align}
where $d=D-1$ denotes the spatial dimension. The Gaussian bridge probability densities $f_{\mathbf{x}(\tau_j)}(\mathbf{r}_{k_j})$ play a central role in this formalism, as they encode the likelihood of the paths passing through successive atomic positions. Each probability density is recursively conditioned on the states of the field at earlier intermediate times, and their product determines the contribution of the field to the interaction potential. The first density, $f_{\mathbf{x}(\tau_2)}(\mathbf{r}_2)$, represents the probability of the path hitting $\mathbf{r}_2$ at time $\tau_2$, given that the path starts and ends at  $\mathbf{r}$ over the total running time $\cT$:
\begin{align}
f_{\mathbf{x}(\tau_2)}(\mathbf{r}_2) = \frac{1}{[2\pi \tau_2 (1-\tau_2/\cT)]^{d/2}}\, e^{-(\mathbf{r}_2 - \mathbf{r})^2 / 2 \tau_2 (1-\tau_2/\cT)}.
\end{align}
Subsequent bridge probabilities depend on the positions of prior intermediate points. Two distinct cases arise based on whether $\tau_3 < \tau_2$ or $\tau_3 > \tau_2$, as the time ordering determines how the field propagates.
If $\tau_3<\tau_2$, the path first reaches $\mathbf{r}_3$ at time $\tau_3$, then proceeds to $\mathbf{r}_2$ at time $\tau_2$, and finally returns to $\mathbf{r}$ at $\cT$. The conditional density is given by
\begin{align}
f_{\mathbf{x}(\tau_3)}(\mathbf{r}_3 \mid \mathbf{x}(\tau_2) = \mathbf{r}_2) &= \frac{1}{[2\pi \tau_3(1-\tau_3/\tau_2)]^{d/2}} e^{-[\mathbf{r}_3 - \mathbf{r}(1-\tau_3/\tau_2)-\mathbf{r}_2\tau_3/\tau_2]^2 / 2\tau_3(1-\tau_3/\tau_2)}.
\label{eq:condition1}
\end{align}
If $\tau_3 > \tau_2$, the path reaches $\mathbf{r}_2$ first, then proceeds to $\mathbf{r}_3$ at time $\tau_3$, before returning to $\mathbf{r}$ at $\cT$. The conditional density is given by
\begin{align}
f_{\mathbf{x}(\tau_3)} (\mathbf{r}_3 \mid \mathbf{x}(\tau_2)&= \mathbf{r}_2) = \frac{1}{[2\pi \tau_{3-} (1-\tau_{3-}/\cT_2)]^{d/2}} e^{-[\mathbf{r}_3 - \mathbf{r}_2(1-\tau_{3-}/\cT_2)-\mathbf{r}\tau_{3-}/\cT_2]^2 / 2 \tau_{3-} (1-\tau_{3-}/\cT_2)},
\label{eq:condition2}
\end{align}
where $\tau_{3-} := \tau_3 - \tau_2$ and $\cT_2 := \cT - \tau_2$. These shifted variables recenter the time intervals around the most recent conditioning point, ensuring that each path segment is described locally by the immediately preceding segment. The overall bridge probability density for a given time ordering is the product of the individual conditional densities. For instance, if $\tau_3 > \tau_2$, the joint probability for the first two bridges is
\begin{align}
    f_{\mathbf{x}(\tau_2)} (\mathbf{r}_2) f_{\mathbf{x}(\tau_3)} &(\mathbf{r}_3 \mid \mathbf{x}(\tau_2)=\mathbf{r}_2) = \frac{e^{-(\mathbf{r}_2 - \mathbf{r})^2 / 2\tau_2 (1-\tau_2/\cT)}}{[2\pi \tau_2 (1-\tau_2/\cT)]^{d/2}} \frac{e^{-[\mathbf{r}_3 - \mathbf{r}_2(1-\tau_{3-}/\cT_2)-\mathbf{r} \tau_{3-}/\cT_2]^2 / 2\tau_{3-} (1-\tau_{3-}/\cT_2)}}{[2\pi \tau_{3-}(1-\tau_{3-}/\cT_2)]^{d/2}} \nonumber \\
    &= \frac{1}{[2\pi \tau_3 (1-\tau_3/\cT)]^{d/2}} e^{-(\mathbf{r}_3 - \mathbf{r})^2/2\tau_3(1-\tau_3/\cT)} \frac{e^{-[\mathbf{r}_2-\mathbf{r}(1-\tau_2/\tau_3)-\mathbf{r}_3 \tau_2/\tau_3]^2/2\tau_2(1-\tau_2/\tau_3)}}{[2\pi \tau_2 (1-\tau_2/\tau_3)]^{d/2}} \nonumber\\
    &= f_{\mathbf{x}(\tau_3)} (\mathbf{r}_3) f_{\mathbf{x}(\tau_2)} (\mathbf{r}_2 \mid \mathbf{x}(\tau_3)=\mathbf{r}_3).
\label{eq:product_f}
\end{align}
A critical feature of these densities is their symmetry under reordering. The equivalence of the prefactors here follows from
\begin{align}
\tau_2 (1 - \tau_2/\cT) \tau_{3-} (1 - \tau_{3-} /\cT_2)&= \tau_2 (\cT - \tau_2) \tau_{3-} (\cT_2-\tau_{3-})/\cT \cT_2 \nonumber\\
&= \tau_2 (\tau_3-\tau_2) (\cT-\tau_3)/\cT \nonumber\\
&= \tau_3 (1-\tau_3/\cT) \tau_2 (1-\tau_2/\tau_3).
\end{align}
This property arises from the fact that if two intermediate points of a bridge are conditioned to have particular states, the order of the conditioning does not matter; that is, $P(A \cap B) = P(A) P(B\! \mid \!A) = P(B) P(A\! \mid \!B)$. It ensures that the path integral remains invariant under reordering of the intermediate points. 

With this time ordering, the recursive nature of the bridge probabilities allows the full path integral for the $k$th order Casimir--Polder potential to be constructed by sequentially integrating over all intermediate positions and times:
\begin{align}
V_{\subCP}^{\subTE(k)} &(\mathbf{r}_1, \ldots, \mathbf{r}_N) = (-1)^{k+1} \frac{(2k - 1)!! \, \hbar c}{2^{k+1} k (2\pi)^{D/2} \epsilon_0^k} \sum_{k_1, \ldots, k_k=1}^N \!\! \alpha_{k_1} \cdots \alpha_{k_k} \int_0^{\infty} \!\! \frac{d\cT}{\cT^{k + D/2}} \, \int_{0 \leq \tau_2 \leq \cdots \leq \tau_k \leq \cT} \prod_{i=2}^k d\tau_i \nonumber \\
& \times f_{\mathbf{x}(\tau_2)} (\mathbf{r}_2) \, f_{\mathbf{x}(\tau_3)} (\mathbf{r}_3 | \mathbf{x}(\tau_2) = \mathbf{r}_2) \cdots f_{\mathbf{x}(\tau_k)} (\mathbf{r}_k | \mathbf{x}(\tau_{k-1}) = \mathbf{r}_{k-1})\, \biggdlangle \langle \epsr \rangle_{\mathbf{x}(\tau)}^{-k - 1/2} \biggdrangle_{\mathbf{x}(\tau) | \mathbf{x}(0) = \mathbf{r}_{k_1}, \mathbf{x}(\tau_j) = \mathbf{r}_{k_j}, \mathbf{x}(\cT) = \mathbf{x}(0)}.
\end{align}
  Note that comparing to Eq.~(\ref{eq:multi_te1}), here we introduced the factor $(k-1)!$ to account for all permutations of the intermediate $\tau_j$. This particular form is convenient as it requires only the conditional density in Eq.~(\ref{eq:condition1}); alternatively, the conditional density in Eq.~(\ref{eq:condition2}) can be used for backward-ordered times.
\subsection{Transverse Magnetic Mode}
The expression of TM potential energy parallels that of the TE case under the duality transformation $\epsr \longleftrightarrow \mur$ and $\alpha_0/\epsilon_0 \longleftrightarrow \beta_0 \mu_0$. From the path-integral point of view, the calculation is complicated by the presence of the nonlinear functions of $\epsr$ in the potential $V\subTM$ as defined in Eq.~(\ref{eq:potential}). If we introduce the atom as the local perturbation and expand the potential to lowest order in $\alpha_0$, we obtain the original potential with an additional atomic interaction term
\begin{align}
V\subTM (\mathbf{x}) &\rightarrow \frac{1}{8} ( \nabla \log \epsr)^2 -\frac{1}{4} \nabla^2 \log \epsr -\frac{\alpha_0}{4\epsilon_0} \big[ (\nabla \log \epsr) \cdot \nabla-\nabla^2 \big] \frac{1}{\epsr} \delta^{D-1} (\mathbf{x}-\mathbf{r}).
\label{eq:vtm_shift}
\end{align}
For an atom in a region of constant $\epsr$, $\nabla \epsilon(\mathbf{r}) = 0$, and the first-order shift in analogy to Eq.~(\ref{eq:vte_final}) is
\begin{align}
V_{\subCP}^{\subTM} (\mathbf{r})&=\frac{\hbar c \alpha_0}{4(2\pi)^{D/2} \epsilon_0} \int_{0}^{\infty}\!\! \frac{d\cT}{\cT^{1+D/2}} \biggdlangle \!\! \left( \frac{1}{\langle \epsilon_{r} \rangle_{\mathbf{x_r}(\tau)}}- \frac{\cT}{2\epsr} \nabla^2 \right)
\langle \epsr \rangle_{\mathbf{x_r}(\tau)}^{-1/2} e^{-\cT \langle V\subTM \rangle_{\mathbf{x_r}(\tau)}} \biggdrangle_{\mathbf{x_r}(\tau)}.
\label{eq:worldlineTM}
\end{align}
Again, this expression lays the foundation of the Monte Carlo evaluation of TM Casimir energies, whose numerical performance is discussed in Ref.~[\onlinecite{Zheng2026}].

The $N=2$ case parallels the TE calculations. By solving for the second-order energy shift and setting $\epsr$ to 1 in vacuum, the two-atom expression for the cross-energy, in the absence of other dielectric objects, is 
\begin{align}
V_{\subCP}^{\subTM}(\mathbf{r}_1, \mathbf{r}_2) &= -\frac{\hbar c \alpha_1 \alpha_2}{8 (2 \pi)^{D-1/2} \epsilon_0^2} \int_0^{\infty} \!\! \frac{d \mathcal{T}}{\mathcal{T}^{2 + D/2}} 
\int_0^{\mathcal{T}} d \tau' \, \left( 3 - \frac{\mathcal{T}}{2} \left( \nabla_1^2 + \nabla_2^2 \right) + \frac{\mathcal{T}^2}{4} \nabla_1^2 \nabla_2^2 \right) \nonumber\\
&\hspace{5mm} \times \frac{1}{\left[ \tau' (1 - \tau'/\mathcal{T}) \right]^{d/2}} e^{-(\mathbf{r}_1 - \mathbf{r}_2)^2/2 \tau' (1 - \tau'/\mathcal{T})}.
\end{align}
Here, the gradient operators $\nabla_i$ differentiate with respect to the atoms' coordinates $\mathbf{r}_i$, while keeping the other coordinate constant. 
For $D=4$, this becomes 
\begin{align}
V_{\subCP}^{\subTM}(r)=-\frac{43 \hbar c \alpha_1 \alpha_2}{8 \pi (4\pi \epsilon_0)^2} \frac{1}{r^7}.
\end{align}
This calculation for the TM component contributes about 93.5\% of the full potential, much larger than the TE's contribution. Again, it differs from the Green tensor's TM contribution, which gives a numerical coefficient of $-73/16$ in Eq.~(\ref{eq:component}). Importantly, the two polarization modes according to worldline calculations add up to the complete result, so the TE/TM decomposition is exact in this case. It is worth noting that both the Green tensor approach and the worldline approach recover the complete interactions. 

Via a similar reasoning in the TE case, a straightforward generalization of the TM Casimir--Polder potential to a discrete system of $N$ atoms at $k$th order is (again, $k\leq N$)
\begin{align}
&V_{\subCP}^{\subTM(k)}(\mathbf{r}_1, \dots, \mathbf{r}_N) = (-1)^{k+1} \frac{\hbar c}{2^{k+1} k! (2 \pi)^{D/2} \epsilon_0^k}\, \sum_{k_1, \dots, k_k = 1}^N \alpha_{k_1} \cdots \alpha_{k_k} \int_0^{\infty} \!\! \frac{d \mathcal{T}}{\mathcal{T}^{k + D/2}} \int_0^{\mathcal{T}} d \tau_2 \cdots d \tau_k \nonumber\\
& \hspace{10mm}\times 
f_{\mathbf{x}(\tau_2)}(\mathbf{r}_2) f_{\mathbf{x}(\tau_3)}(\mathbf{r}_3) \cdots f_{\mathbf{x}(\tau_k)}(\mathbf{r}_k) 
\biggdlangle \prod_{j=1}^k \left[ \frac{2j - 1}{\langle \epsr \rangle_{\mathbf{x}(\tau)}} - \frac{\mathcal{T}}{2 \epsr(\mathbf{r}_{k_j})} \nabla_{k_j}^2 \right] 
\frac{e^{-\cT \langle V\subTM \rangle}}{\langle \epsr \rangle^{1/2}_{\mathbf{x}(\tau)}} \biggdrangle_{\mathbf{x}(\tau) | \mathbf{x}(0) = \mathbf{r}_{k_1}, \mathbf{x}(\tau_j) = \mathbf{r}_{k_j}, \mathbf{x}(\cT) = \mathbf{x}(0)},
\end{align}
After a time-ordered pinning of the paths, we obtain 
\begin{align}
&V_{\subCP}^{\subTM(k)}(\mathbf{r}_1, \dots, \mathbf{r}_N) = (-1)^{k+1} \frac{\hbar c}{2^{k+1} k (2 \pi)^{D/2} \epsilon_0^k} \sum_{k_1, \dots, k_k = 1}^N \alpha_{k_1} \cdots \alpha_{k_k} \int_0^{\infty} \!\! \frac{d \mathcal{T}}{\mathcal{T}^{k + D/2}} \, \int_{0 \leq \tau_2 \leq \cdots \leq \tau_k \leq \mathcal{T}} \prod_{i=2}^k d\tau_i \nonumber\\
& \hspace{10mm}\times f_{\mathbf{x}(\tau_2)} (\mathbf{r}_2) f_{\mathbf{x}(\tau_3)}(\mathbf{r}_3 \mid \mathbf{x}(\tau_2) = \mathbf{r}_2) \cdots f_{\mathbf{x}(\tau_k)}(\mathbf{r}_k \mid \mathbf{x}(\tau_{k-1}) = \mathbf{r}_{k-1})
\nonumber\\
&\hspace{10mm}\times 
\biggdlangle \prod_{j=1}^k \left[ \frac{2j - 1}{\langle \epsr \rangle_{\mathbf{x}(\tau)}} - \frac{\mathcal{T}}{2 \epsr(\mathbf{r}_{k_j})} \nabla_{k_j}^2 \right]
\frac{e^{-\cT \langle V\subTM \rangle}}{\langle \epsr \rangle^{1/2}_{\mathbf{x}(\tau)}} \biggdrangle_{\mathbf{x}(\tau) | \mathbf{x}(0) = \mathbf{r}_{k_1}, \mathbf{x}(\tau_j) = \mathbf{r}_{k_j}, \mathbf{x}(\cT) = \mathbf{x}(0)}.
\end{align}
The ordering of the operator product does not matter here due to the commutativity of partial derivatives.

\subsection{Analytic Results for N = 3}
\label{sec:analyticN=3}
To further investigate the validity of the scalar decomposition, we apply these general expressions of Casimir--Polder interactions in the three-body case. For $N=3$, we need one more intermediate time $\tau_3$ to pin the paths at $\mathbf{r}_3$. Starting with the TE component, the three-atom expression for cross-energy only is given by
\begin{align}
V_{\subCP}^{\subTE(3)} (\mathbf{r}_1,\mathbf{r}_2,\mathbf{r}_3)&=\frac{15 \hbar c \alpha_1 \alpha_2 \alpha_3}{48(2\pi)^{D/2} \epsilon_0^3} \int_0^{\infty} \!\! \frac{d\cT}{\cT^{3+D/2}} 
\int_{0 \leq \tau_2 \leq \tau_3 \leq \cT} \!\!d\tau_2 \, d\tau_3 \, 
f_{\mathbf{x}(\tau_2)}(\mathbf{r}_{2}) f_{\mathbf{x}(\tau_3)}(\mathbf{r}_{3}|\mathbf{x}(\tau_2)=\mathbf{r}_{2}) \nonumber\\
&=\frac{15 \hbar c \alpha_1 \alpha_2 \alpha_3}{48(2\pi)^{D/2} \epsilon_0^3} \int_0^{\infty} \!\! \frac{d\cT}{\cT^{3+D/2}}\, \int_0^{\cT} \!\! d\tau_2 \, \int_{\tau_2}^{\cT} \!\! d\tau_3 \, \frac{1}{[2\pi \tau_3 (1-\tau_3/\cT)]^{d/2}} e^{-(\mathbf{r}_3-\mathbf{r}_1)^2/2\tau_3(1-\tau_3/\cT)} \nonumber\\
&\hspace{10mm}\times \frac{e^{-[\mathbf{r}_2-\mathbf{r}_1(1-\tau_2/\tau_3)-\mathbf{r}_3 \tau_2/\tau_3]^2/2\tau_2 (1-\tau_2/\tau_3)}}{[2\pi \tau_2 (1-\tau_2/\tau_3)]^{d/2}}.
\end{align}
Here, we time-ordered the crossing time through $\mathbf{r}_2$ to occur before that through $\mathbf{r}_3$. For this reason, a path that originates from $\mathbf{r}_1$ running in time $\cT$ can touch the second atom at all possible $\tau_2$ between 0 and $\cT$. Once $\tau_2$ is fixed, the path can only cross the third atom between $\tau_2 < \tau_3 < \cT$. This choice of time-ordering is reflected in the integration boundaries in the expression.

The TM potential for three atoms is complicated by derivative operators. It takes the form 
\begin{align}
V_{\subCP}^{\subTM(3)}(\mathbf{r}_1, \mathbf{r}_2, \mathbf{r}_3)&=\frac{\hbar c \alpha_1 \alpha_2 \alpha_3}{48(2\pi)^{D/2} \epsilon_0^3} \int_0^{\infty} \!\! \frac{d\cT}{\cT^{3+D/2}} \, \int_0^\cT \!\! d\tau_2 
\int_{\tau_2}^\cT \!\! d\tau_3 \, \bigg [ 15-\frac{3\cT}{2} \nabla_1^2 -\frac{3\cT}{2} \nabla_2^2 -\frac{3\cT}{2} \nabla_3^2 +\frac{\cT^2}{4} \nabla_1^2 \nabla_2^2 \nonumber\\
&+\frac{\cT^2}{4} \nabla_1^2 \nabla_3^2 +\frac{\cT^2}{4} \nabla_2^2 \nabla_3^2-\frac{\cT^3}{8} \nabla_1^2 \nabla_2^2 \nabla_3^2 \bigg ] \frac{1}{[2\pi \tau_3 (1-\tau_3/\cT)]^{d/2}}e^{-(\mathbf{r}_3-\mathbf{r}_1)^2/2\tau_3(1-\tau_3/\cT)} \nonumber\\
&\times \frac{e^{-[\mathbf{r}_2-\mathbf{r}_1(1-\tau_2/\tau_3)-\mathbf{r}_3 \tau_2/\tau_3]^2/2\tau_2 (1-\tau_2/\tau_3)}}{[2\pi \tau_2 (1-\tau_2/\tau_3)]^{d/2}}. 
\end{align}
At this point, the $\mathbf{r}_i$ still remain unspecified. We examine two configurations: three atoms on a line ($R/2$ spacing) and on an equilateral triangle of side length $R$. 
For the collinear case, the breakdown of the scalar approximation is substantial. The worldline scalar sum gives 
\begin{align}
\Delta E \approx +1074 \frac{\hbar c \alpha_1 \alpha_2 \alpha_3}{\pi} \frac{1}{(4\pi \epsilon_0)^3 R^{10}}, 
\end{align} 
where the Green tensor result in Eq.~(\ref{eq:colinear_correct}) gives a coefficient of $-186$.  This is a large discrepancy in both magnitude and sign, indicating this configuration does not respect the scalar decomposition.
Similarly, for the triangular case, the two scalar modes sum to an energy shift of
\begin{align}
\Delta E \approx -5.7 \frac{\hbar c \alpha_1 \alpha_2 \alpha_3}{\pi} \frac{1}{(4\pi \epsilon_0)^3 R^{10}}. 
\end{align}
This does not agree with the Green tensor's scattering results from Eq.~(\ref{eq:triangle_correct}), where the coefficient is $+5.2$.

\begin{figure}[tb]
\begin{center}
\includegraphics[width=\textwidth]{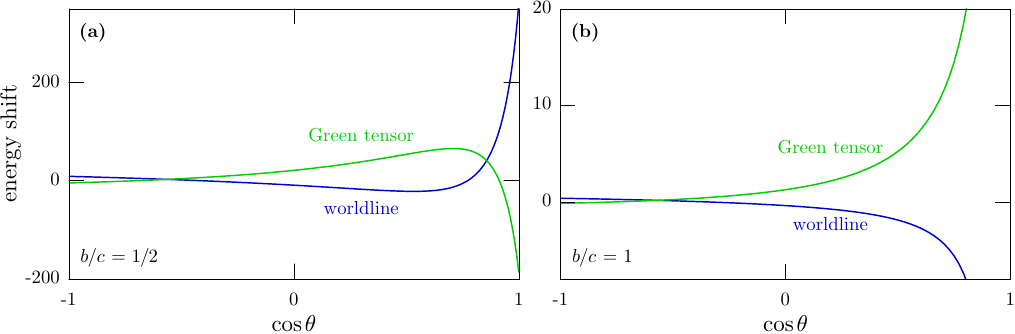}
\end{center}
\caption{
The Casimir--Polder coefficients for the three-body interaction obtained from the worldline scalar sum and the Green tensor method, as a function of the cosine of the angle between the displacement vectors.  The comparison for $b/c=0.5$ is shown
in (a), where the collinear configuration corresponds to $\cos \theta=1$.  The case $b/c=1$ is shown in (b), where
the equilateral-triangle configuration corresponds to $\cos\theta=0.5$.
In both cases, the energy shift is plotted in units of
$\hbar c\alpha_1\alpha_2\alpha_3/(4\pi\epsilon_0)^3\pi R^{10}$.
}
\label{fig:data}
\end{figure}

To investigate the geometry dependence of the breakdown of the scalar worldline sum, we consider three atoms \( A \), \( B \), and \( C \), placed in a two-dimensional plane with absolute positions defined as
\begin{align}
\mathbf{r}_A &= (0, 0, 0), \nonumber \\
\mathbf{r}_B &= (b, 0, 0), \nonumber \\
\mathbf{r}_C &= (c \cos\theta, c \sin\theta, 0) \label{eq:positions}
\end{align}
in terms of the parameters $b$ and $c$.
For visualization, we evaluate the numerical coefficients for the interaction potential using both the worldline method and the exact Green-tensor method vs.\ the cosine of the orientation angle
(similar to Ref.~[\onlinecite{Power1985}]).
By this convention, $\cos \theta=0.5$ with $b/c=1$ corresponds to an equilateral triangle and $\cos \theta=1$ with $b/c=0.5$ corresponds to a linear configuration with $C$ to the right of $A$.
The comparison between the two computational methods is shown for two fixed values of $b/c$ in Fig.~\ref{fig:data}a ($b/c=1/2$) and Fig.~\ref{fig:data}b ($b/c=1$). We observe that the shapes of their curves differ non-trivially across $\cos \theta$, further highlighting the discrepancy between the worldline and Green-tensor methods.

\section{Discussion}
The Green-tensor and worldline formalisms clearly
give different results in the three-body case, although
they in principle use the same TE/TM decomposition.
Note that in Eq.~(\ref{eq:vtensor}), it is the product of 
Green tensors that appears, and each Green tensor has its own domain of
integration. The same thing happens in the three-body case,
Eq.~(\ref{eq:3bd-pot}).
Thus, the calculation in the Green-tensor formalism
involves direct TE and TM contributions
as well as cross-coupling contributions. 
By contrast, the worldline method within the scalar approximation
makes the TE/TM decomposition at the level of the scalar
wave equations.
It is for this reason that the decompositions are not equivalent.
In the two-body case, the contributions from the different
polarization components do not match, but fortuitously, the
sum of all the components does match between the two methods.
However, the agreement breaks down in the three-body case. The scalar worldline approach yields total interaction potentials that differ noticeably in magnitude as well as in sign compared to the exact results, across a variety of three-body geometric configurations.
This calculation suggests that strong polarization coupling exists in multiple-body systems, and motivates the extension of the worldline formalism beyond the scalar approximation to fully account for electromagnetism.

\section{Summary}
In this study, we explore the limits of the scalar approximation within the worldline path-integral formulation in modeling multiple-atom Casimir--Polder interactions. 
In particular, we derived the Casimir--Polder potential within 
the scalar worldline formalism for $N$ atoms.
For two-atom systems, the scalar method yields the expected energies.  This is analogous to the image method,\cite{Mackrory2016} where the ``image'' is effectively replaced by the second atom. However, when extending the approach to three-atom systems, we identify a regime in which the scalar approximation disagrees substantially with the correct result. Our findings emphasize the important role of polarization mixing in accurately describing multiple-atom Casimir--Polder interactions.

\appendix*
\section{N=3 Potential}
\label{app:3pot}
Suppose three atoms are located along the $z$-axis, with positions $z_1,z_2,z_3$ such that $|z_3-z_2|=|z_2-z_1|=R/2$. This case is a straightforward generalization of the $N=2$ case. The potential in terms of the Green tensors is given by Eq.~(\ref{eq:3bd-pot}).
To evaluate the first term in the trace, this leads to integrals over all possible $k$'s:
\begin{align}
V\subCP^{(3)}&=-\frac{\hbar \alpha_1 \alpha_2 \alpha_3}{2\pi (8\pi^2 \epsilon_0)^3} \int_0^{\infty} \!\! ds \, \int_{-\infty}^{\infty} \!\! dk_x dk_y dk_x' dk_y' dk_x'' dk_y'' \frac{-s^6/c^6}{\kappa(s,k_x,k_y)\kappa(s,k_x',k_y')\kappa(s,k_x'',k_y'')} \nonumber\\
&\hspace{5mm} \times \delta_{ab} \delta_{cd} \delta_{ef} (e_b e_c+h_b h_c)(e'_d e'_e+h'_d h'_e)(e''_f e''_a+h''_f h''_a) e^{-\kappa(s,k_x,k_y) z} e^{-\kappa(s,k_x',k_y') z'} e^{-\kappa(s,k_x'',k_y'') z''}.
\end{align}
Expanding out the tensor part of this expression, we obtain 8 terms given by
\begin{align}
&\delta_{ab} \delta_{cd} \delta_{ef} (\hat{e}_b \hat{e}_c+\hat{h}_b \hat{h}_c)(\hat{e}'_d \hat{e}'_e+\hat{h}'_d \hat{h}'_e)(\hat{e}''_f \hat{e}''_a+\hat{h}''_f \hat{h}''_a)\nonumber\\
&=(\hat{e}_a \hat{e}_d+\hat{h}_a \hat{h}_d)(\hat{e}'_d \hat{e}'_f+\hat{h}'_d \hat{h}'_f)(\hat{e}''_f \hat{e}''_a+\hat{h}''_f \hat{h}''_a)\nonumber\\
&=\mathrm{Tr}(\mathbf{A}_e \mathbf{A}_e' \mathbf{A}_e'') + \mathrm{Tr}(\mathbf{A}_e' \mathbf{A}_e'' \mathbf{A}_h)+\mathrm{Tr}(\mathbf{A}_e \mathbf{A}_e'' \mathbf{A}_h') \nonumber\\
&\hspace{5mm}+\mathrm{Tr}(\mathbf{A}_e'' \mathbf{A}_h \mathbf{A}_h')+\mathrm{Tr}(\mathbf{A}_e \mathbf{A}_e' \mathbf{A}_h'')+\mathrm{Tr}(\mathbf{A}_e' \mathbf{A}_h \mathbf{A}_h'') \nonumber\\
&\hspace{5mm}+\mathrm{Tr}(\mathbf{A}_e \mathbf{A}_h' \mathbf{A}_h'')+\mathrm{Tr}(\mathbf{A}_h \mathbf{A}_h' \mathbf{A}_h'').
\end{align}
Carrying out the traces, we can write the TE and TM components as
\begin{align}
\mathrm{Tr}(\mathbf{A}_e \mathbf{A}_e' \mathbf{A}_e'')&=\frac{(k_x k_x'+k_y k_y') (k_x k_x''+k_y k_y'') (k_x' k_x''+k_y' k_y'')}{k_\rho^2 k_\rho'^2 k_\rho''^2} \nonumber\\
\mathrm{Tr}(\mathbf{A}_h \mathbf{A}_h' \mathbf{A}_h'') &=\frac{k_\rho^2 k_\rho'^2+k_x k_x' k_z k_z'+k_y k_y' k_z k_z'}{k^2 k'^2 k''^2 k_\rho^2 k_\rho'^2 k_\rho''^2} \nonumber\\
&\hspace{5mm}\times (k_\rho^2 k_\rho''^2+k_x k_x'' k_z k_z''+k_y k_y'' k_z k_z'') \nonumber\\
&\hspace{5mm}\times (k_\rho'^2 k_\rho''^2+k_x' k_x'' k_z' k_z''+k_y' k_y'' k_z' k_z'').
\end{align}
We start with the TE term, which evaluates to
\begin{align}
V_{\subTE}^{(3)}(R)&=\frac{\hbar \alpha_1 \alpha_2 \alpha_3}{2\pi} (\frac{1}{8\pi^2 \epsilon_0})^3 (2\pi^3) \int_0^{\infty} \!\!ds\, \int_{0}^{\infty} \!\! dk_\rho dk_\rho' dk_\rho'' \nonumber\\
&\hspace{5mm}\times \frac{s^6 k_\rho k_\rho' k_\rho''}{c^6 \kappa(s,k_x,k_y)\kappa(s,k_x',k_y')\kappa(s,k_x'',k_y'')} \, e^{-\kappa(s,k_x,k_y) z} e^{-\kappa(s,k_x',k_y') z'} e^{-\kappa(s,k_x'',k_y'') z''} |_{z=z'=R/2, z''=R}\nonumber\\
&=\frac{45\hbar c \alpha_1 \alpha_2 \alpha_3}{16\pi} \frac{1}{(4\pi \epsilon_0)^3 R^{10}}.
\end{align} 
Recall the worldline TE result gives $45/2 = +22.5$, whereas the coefficient from the current method is about $45/16=+2.8125$. Note that the magnitude is significantly smaller than the worldline TE result. 
\par
Similarly, the TM contribution from the corresponding terms in the Green tensors is 
\begin{align}
V_{\subTM}^{(3)}(R)&=-\frac{\hbar \alpha_1 \alpha_2 \alpha_3}{2\pi} (\frac{1}{8\pi^2 \epsilon_0})^3 \int_0^{\infty} \!\! ds\, \int_{0}^{\infty} \!\! dk_\rho dk_\rho' dk_\rho'' \nonumber\\
&\hspace{5mm}\times \frac{k_\rho k_\rho' k_\rho'' [(2\pi)^3 k_\rho^2 k_\rho'^2 k_\rho''^2+2\pi^3 k_z^2 k_z'^2 k_z''^2]}{\kappa(s,k_x,k_y)\kappa(s,k_x',k_y')\kappa(s,k_x'',k_y'')} \, e^{-\kappa(s,k_x,k_y) z} e^{-\kappa(s,k_x',k_y') z'} e^{-\kappa(s,k_x'',k_y'') z''} |_{z=z'=R/2, z''=R} \nonumber\\
&=-\frac{3297\hbar c \alpha_1 \alpha_2 \alpha_3}{16\pi} \frac{1}{(4\pi \epsilon_0)^3 R^{10}}.
\end{align} 
The numerical coefficient for the pure TM term is about $-3297/16 \approx -206.063$. The worldline method gives the TM coefficient as +1051.5, however.
Finally, we consider the TE/TM mixing terms. These terms arise from traces involving combinations of TE ($\mathbf{A}_e$) and TM ($\mathbf{A}_h$) contributions. The six distinct traces can be evaluated as:
\begin{align}
\mathrm{Tr}(\mathbf{A}_e' \mathbf{A}_e'' \mathbf{A}_h)& \rightarrow 2\pi^3 \frac{k_z^2}{k^2} \nonumber\\
\mathrm{Tr}(\mathbf{A}_e \mathbf{A}_e'' \mathbf{A}_h')& \rightarrow 2\pi^3 \frac{k_z'^2}{k'^2} \nonumber\\
\mathrm{Tr}(\mathbf{A}_e \mathbf{A}_e' \mathbf{A}_h'')& \rightarrow 2\pi^3 \frac{k_z''^2}{k''^2} \nonumber\\
\mathrm{Tr}(\mathbf{A}_e'' \mathbf{A}_h \mathbf{A}_h')& \rightarrow 2\pi^3 \frac{k_z^2 k_z'^2}{k^2 k'^2} \nonumber\\
\mathrm{Tr}(\mathbf{A}_e' \mathbf{A}_h \mathbf{A}_h'')& \rightarrow 2\pi^3 \frac{k_z^2 k_z''^2}{k^2 k''^2} \nonumber\\
\mathrm{Tr}(\mathbf{A}_e \mathbf{A}_h' \mathbf{A}_h'')& \rightarrow 2\pi^3 \frac{k_z'^2 k_z''^2}{k'^2 k''^2}.
\end{align}
The first three are mixed TE-dominated terms and the last three are mixed TM-dominated terms. For the former type, the first term gives
\begin{align}
V^{\subTE/\subTM}_{1}(R)&=-\frac{\hbar \alpha_1 \alpha_2 \alpha_3}{2\pi} (\frac{1}{8\pi^2 \epsilon_0})^3 (2\pi^3) \int_0^{\infty} \!\! ds \, \int_{0}^{\infty} \!\! dk_\rho \nonumber\\
&\hspace{5mm}\times \int_{0}^{\infty} \!\! dk_\rho' \int_{0}^{\infty} \!\! dk_\rho'' \frac{-k_\rho k_\rho' k_\rho'' s^6/c^6}{\kappa(s,k_x,k_y)\kappa(s,k_x',k_y')\kappa(s,k_x'',k_y'')} \frac{k_z^2}{k^2} \nonumber\\
&\hspace{5mm}\times e^{-\kappa(s,k_x,k_y) z} e^{-\kappa(s,k_x',k_y') z'} e^{-\kappa(s,k_x'',k_y'') z''} |_{z=z'=\frac{R}{2}, z''=R} \nonumber\\
&=\frac{153 \hbar c \alpha_1 \alpha_2 \alpha_3}{16 \pi} \frac{1}{(4\pi \epsilon_0)^3 R^{10}}.
\end{align} 
The coefficient of this TE/TM mixing potential is about $153/16 \approx +9.5625$, with a positive sign. The second term should produce an identical result since $k_z \leftrightarrow k_z'$ and thus $z \leftrightarrow z'$, given $z=z'$. However, the third term involving $k_z''$ may not be identical, since $z''$ takes a different value from the other two. After calculation, such term contributes a coefficient of +5.4375.

Finally, we evaluate the traces of the second type of the polarization mixing matrices, the fourth cross-term involving $k_z$ and $k_z'$ gives 
\begin{align}
V^{\subTE/\subTM}_{2}(R)&=-\frac{\hbar \alpha_1 \alpha_2 \alpha_3}{2\pi} (\frac{1}{8\pi^2 \epsilon_0})^3 (2\pi^3) \int_0^{\infty}\!\! ds\, \int_{0}^{\infty} \!\! dk_\rho \nonumber\\
&\hspace{5mm}\times \int_0^{\infty}\!\! dk_\rho' \,\int_0^{\infty}\!\! dk_\rho''\, \frac{-s^6 k_\rho k_\rho' k_\rho'' k_z^2 k_z'^2/(k^2 k'^2)}{\kappa(s,k_x,k_y)\kappa(s,k_x',k_y')\kappa(s,k_x'',k_y'') c^6} \nonumber\\
&\hspace{5mm}\times e^{-\kappa(s,k_x,k_y) z} e^{-\kappa(s,k_x',k_y') z'} e^{-\kappa(s,k_x'',k_y'') z''} |_{z=z'=\frac{R}{2}, z''=R} \nonumber\\
&=\frac{677 \hbar c \alpha_1 \alpha_2 \alpha_3}{16 \pi} \frac{1}{(4\pi \epsilon_0)^3 R^{10}}.
\end{align} 
This cross-term produces a coefficient of about $677/16 \approx +42.3125$.
The fifth term should be identical to the sixth term. With $z' \leftrightarrow z''$, we obtain a coefficient of +21.6875.

Summing all the contributions together, the Casimir--Polder interaction from these polarizations is given by 
\begin{align}
V_{\subTE}^{(3)} + V_{\subTM}^{(3)} + \sum V^{(3)}_{\subTE/\subTM} =-93 \frac{\hbar c \alpha_1 \alpha_2 \alpha_3}{\pi} \frac{1}{(4\pi \epsilon_0)^3 R^{10}}.
\label{eq:3bdfinal}
\end{align} 
Recall that in Eq.~(\ref{eq:3bd-pot}), we have two terms in the trace that should give identical results. Thus, the total  Casimir--Polder potential for three collinear atoms is just double of Eq.~(\ref{eq:3bdfinal}),
\begin{align}
V_{\subCP}^{(3)}(R) = -186 \frac{\hbar c \alpha_1 \alpha_2 \alpha_3}{\pi} \frac{1}{(4\pi \epsilon_0)^3 R^{10}},
\end{align} 
which is consistent with the three-body collinear results from Ref.~[\onlinecite{Power1985}].

In this calculation, the Green tensor contraction is completely general. The only geometric-specific information resides in the evaluation of the spatial coordinates at the end. For the collinear configuration shown above, the system is one-dimensional and only $\{z,z',z''\}$ need to be specified. For three atoms located at the vertices of an equilateral triangle, the calculations proceed in the same manner after we identify the coordinates $\{(x_1,y_1),(x_2,y_2),(x_3,y_3) \}$ explicitly.
\nocite{*}
%

\end{document}